\documentclass[letter]{article}
\usepackage[caption=false]{subfig}
\usepackage{pgfplots,tikz}

\usepackage{INTERSPEECH2020}

\title{GEV Beamforming Supported by DOA-based Masks\\ Generated on Pairs of Microphones}
\name{Fran\c{c}ois Grondin, Jean-Samuel Lauzon, Jonathan Vincent, Fran\c{c}ois Michaud}
\address{
  Universit\'e de Sherbrooke, Sherbrooke (Qu\'ebec), Canada}
\email{\{francois.grondin2,jean-samuel.lauzon,jonathan.vincent2,francois.michaud\}@usherbrooke.ca}

\begin{document}

\maketitle
\begin{abstract}
Distant speech processing is a challenging task, especially when dealing with the cocktail party effect.
Sound source separation is thus often required as a preprocessing step prior to speech recognition to improve the signal to distortion ratio (SDR).
Recently, a combination of beamforming and speech separation networks have been proposed to improve the target source quality in the direction of arrival of interest.
However, with this type of approach, the neural network needs to be trained in advance for a specific microphone array geometry, which limits versatility when adding/removing microphones, or changing the shape of the array.
The solution presented in this paper is to train a neural network on pairs of microphones with different spacing and acoustic environmental conditions, and then use this network to estimate a time-frequency mask from all the pairs of microphones forming the array with an arbitrary shape.
Using this mask, the target and noise covariance matrices can be estimated, and then used to perform generalized eigenvalue (GEV) beamforming.
Results show that the proposed approach improves the SDR from 4.78 dB to 7.69 dB on average, for various microphone array geometries that correspond to commercially available hardware.
\end{abstract}

\section{Introduction}

Distant speech processing is a challenging task, as the target speech signal is often corrupted by additive noise and reverberation from the environment \cite{tang2018study}.
Moreover, robust speech recognition often relies on sound source separation when dealing with the cocktail party effect.
Speech separation methods can be divided in two main categories: blind speech separation and informed speech separation.
Blind speech separation relies strictly on the mixture spectrogram to restore the individual sources, whereas informed speech separation uses additional information such as video, direction of arrival and speaker features.

Blind speech separation is particularly challenging as it needs to solve the permutation ambiguity.
In fact, the order of the separated signals may differ from the order of the labels, which makes supervised learning difficult. 
To solve this issue, deep clustering (DC) uses contrastive embedding vectors and unsupervised clustering using k-means \cite{hershey2016deep,luo2018speaker,wang2018deep,isik2016single,wang2018multi}.
Alternatively, permutation invariant training (PIT) aims to find all possible permutations during training and keep the optimal one \cite{yu2017permutation,qian2018single,yoshioka2018multi}.
These methods aim to separate all sources in the mixture, though sometimes only a specific target source matters.
In the latter case, using an objective function that emphasizes only on the target source leads to better performance \cite{chen2017cracking}.

Informed speech separation relies on additional information to perform separation.
For instance, SpeakerBeam uses the speaker identification features to extract a specific speaker from a mixture \cite{vzmolikova2019speakerbeam}.
When the video is also available, it is possible to solve the permutation issue by combining the audio signal and the motion of the lips \cite{ephrat2018looking,afouras2018conversation,petridis2018end}.
Moreover, when dealing with multiple channels, it is often common to use the direction of arrival (DOA) of sound to solve permutation \cite{chen2018efficient}.
Beamforming is thus a special case of informed speech separation, as it exploits the spatial information to reconstruct the target source.
This paper focuses on a beamforming approach that exploits the target source DOA information.

Delay and Sum (DS) and Minimum Variance Distortionless Response (MVDR) beamformers \cite{habets2009new,erdogan2016improved,xiao2017time} improve the signal to noise ratio (SNR) of the target sound source, but relies on DOA of sound derived from the anechoic model for sound propagation, which often differs from the actual condition in a reverberant environment.
On the otherhand, Generalized eigenvalue decomposition (GEV) beamforming maximizes the SNR using only the target and interfering signals covariance matrices \cite{warsitz2007blind}.
Heymann et al. \cite{heymann2015blstm,heymann2016neural,heymann2017beamnet} show that these covariance matrices can be estimated with a bi-directional Long Short-Term Memory (BLSTM) network trained on noisy speech, and that blind analytic normalization (BAN) gain minimizes non-linear distortion for the separated signal.
To be effective, this approach assumes that the interfering sounds differ from speech, which is a major limitation when dealing with the cocktail party effect.
Chen et al. \cite{chen2018multi} propose to use the DOA of sound to estimate a time-frequency mask of a target source with a neural network, and then use this mask to compute the target and noise covariance matrices.
This approach performs well but has one major drawback: the geometry of the microphone array needs to be known prior to training the neural network, which impacts considerably the versability of the system when dealing with microphone arrays of arbitrary shapes.
Maldonado et al. \cite{maldonado2020lightweight} present a solution to deal with the arbitrary shape, but the time-frequency mask obtained from the microphone array DOA is essentially applied to a single channel to extract the target spectrum and no further beamforming is used during separation.
Liu et al. \cite{liu2018neural} also propose to estimate a time-frequency mask based on the cross-spectrum between two microphones and the target time difference of arrival (TDOA), but their approach is limited to two microphones and the spacing between the microphones is fixed.

The method presented in this paper, called SteerNet, relies on a neural network trained on pairs of microphones with different spacing.
This network generates a time-frequency soft mask for each pair of microphones for a set of target TDOAs, obtained from the DOA of the target source and the array geometry.
These masks are combined and used to compute target and noise covariance matrices and to perform GEV beamforming.
This method is appealing as it makes the best use of GEV beamforming using DOA to solve permutation, while being able to generalize to microphone arrays of arbitrary shapes.

\section{SteerNet}
\label{sec:proposed_method}

\begin{figure*}[!ht]
    \centering
    \includegraphics[width=\linewidth]{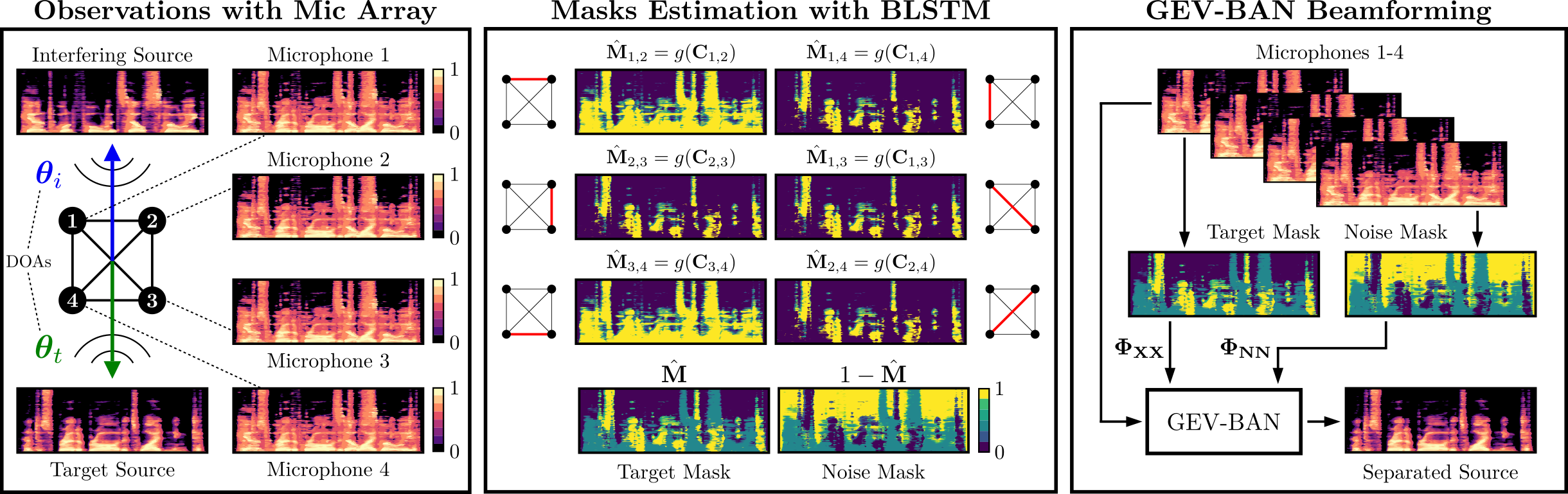}
    \vspace{-15pt}
    \caption{Overview of SteerNet. In this example, the TDOAs of the target and interference are identical for the pairs (1,2) and (3,4), and thus the mask captures both sources, whereas the mask discriminates the target from the interfering source with other pairs. These masks are then used to compute the covariance matrices for the target ($\bm{\Phi}_{\mathbf{XX}}$) and noise ($\bm{\Phi}_{\mathbf{NN}}$) signals, and then GEV-BAN beamforming produces the separated source.}
    \label{fig:overview}
\end{figure*}

Figure \ref{fig:overview} shows the SteerNet method to separate a target speech source using a microphone array with an arbitrary geometry.
In this scenario, there are two speech sources, the target and interference, and it is assumed that these sources have different DOAs.
SteerNet assumes that the DOA of the target speech is available and is obtained using sound source localization methods \cite{grondin2019svd,grondin2019multiple,grondin2019lightweight}, or using a visual cue when both optical and acoustic images are properly aligned \cite{grondin2020audio}.

Using the target DOA, the idea is to generate a time-frequency mask using a neural network to capture the target source components.
To deal with arbitrary geometries, the method breaks down the shape of the array in pairs of microphones, and use the TDOAs between microphones to generate multiple masks.
Masks that put emphasis on the target source can be estimated when the target and interference TDOAs are different as the permutation is easily solved.
This is the case with most pairs of microphones, yet some of them can have similar TDOAs.
When both TDOAs are similar, SteerNet generates a mask that capture both sources, as permutation cannot be solved spatially.
The overall target source mask is obtained by summing all the estimated masks amongst all pairs of microphones.
This leads to a target mask that emphasizes the time-frequency region dominated by the target source due to the pairs of microphones that allow discrimination between sources.
The noise mask is obtained as the complement of the target.
The approach finally uses GEV-BAN beamforming, which relies on the target and noise covariance matrices (denoted as $\bm{\Phi}_{\mathbf{XX}}$ and $\bm{\Phi}_{\mathbf{NN}}$, respectively) obtained from the estimated masks.

\subsection{Oracle pairwise ratio mask}

Let's define the DOAs of the target and interference as $\bm{\theta}_t \in \mathcal{S}^2$ and $\bm{\theta}_i \in \mathcal{S}^2$ respectively, where $\mathcal{S}^2 = \{\mathbf{x} \in \mathbb{R}^3: \lVert \mathbf{x} \rVert_2 = 1\}$, holds unit vectors and $\|\dots\|_2$ stands for the Euclidean norm.
Let's also define the set of indexes of microphone pairs as $\mathcal{Q} = \{(x,y) \in \mathcal{D}^2: x < y \}$, where the set $\mathcal{D} = \{1, 2, \dots, D\}$ contains the microphone indexes, and $D$ stands for the number of microphones.
The TDOA for microphones $(u,v) \in \mathcal{Q}$ corresponds to the following expression:
\begin{equation}
    \tau_{u,v} = \frac{f_S}{c}(\mathbf{r}_u - \mathbf{r}_v) \cdot \bm{\theta}_t,
\end{equation}
where $\mathbf{r}_u$, $\mathbf{r}_v$, $f_S$ and $c$ stand for the positions of microphones $u$ and $v$ (in m), the sample rate (in sample/sec) and speed of sound (in m/sec), respectively.
The steering vector in the direction of the target for a pair of microphones is defined as:
\begin{equation}
    A_{u,v}(t,f) = \exp{\left(j\frac{2\pi f \tau_{u,v}}{N}\right)},
\end{equation}
where $N$ stands for the number of samples per frame in the Short Time Fourier Tranform (STFT), $t$ the frame index, and $f$ the frequency bin index.

Similarly, the difference between TDOAs for a given microphone pair associated to the target and interfering speech sources is estimated as: 
\begin{equation}
    \Delta\tau_{u,v} = \frac{f_S}{c} |(\bm{\theta}_t-\bm{\theta}_i)\cdot(\mathbf{r}_u-\mathbf{r}_v)|.
\end{equation}

The steering vector aims to cancel the phase difference of the target source in the cross-spectrum between microphones $u$ and $v$: 
\begin{equation}
    Y_{u,v}(t,f) = A_{u,v}(t,f) Y_{u}(t,f) Y_{v}(t,f)^*,
\end{equation}
where the expression $\{\dots\}^*$ stands for the complex conjugate, and $Y_u(t,f)$ and $Y_v(t,f)$ stand for the spectra of microphones $u$ and $v$, respectively.
The gain $G_{u,v}$ is defined as a function of $\Delta\tau_{u,v}$, where the goal is to ensure it goes to a value of $1$ when both TDOAs are similar, and goes to zero when they are different.
To smooth the transition and control the sharpness, a sigmoid function is used ($\alpha$ is the steepness and $\beta$ the offset):
\begin{equation}
    G_{u,v} = \frac{\exp{\{-\alpha(\Delta\tau_{u,v}-\beta)\}}}{1+\exp{\{-\alpha(\Delta\tau_{u,v}-\beta)\}}}.
    \label{eq:G}
\end{equation}

This gain is then used to generate the ideal ratio mask for microphones $u$ and $v$:
\begin{equation}
    M_{\bar{u},v}(t,f) = \frac{|S_u(t,f)|^2 + G_{u,v}|I_u(t,f)|^2}{|S_u(t,f)|^2 + |I_u(t,f)|^2 + |B_u(t,f)|^2}
    \label{eq:M_ubarv},
\end{equation}
\begin{equation}
    M_{u,\bar{v}}(t,f) = \frac{|S_v(t,f)|^2 + G_{u,v}|I_v(t,f)|^2}{|S_v(t,f)|^2 + |I_v(t,f)|^2 + |B_v(t,f)|^2}
    \label{eq:M_uvbar}.
\end{equation}

When both the target and interference share a similar TDOA, the gain goes to one and the oracle mask captures both the target ($S_u(t,f)$ and $S_v(t,f)$) and interference ($I_u(t,f)$ and $I_v(t,f)$), and rejects the diffuse background noise ($B_u(t,f)$ and $B_v(t,f)$).
On the otherhand, when discrimination between the target and interference is possible due to different TDOAs, the gain goes to $0$ and the oracle mask captures only the target source.
Finally, the mask for a pair of microphones $(u,v)$ (denoted as $M_{u,v}(t,f)$) is obtained as follows:
\begin{equation}
    M_{u,v}(t,f) = M_{\bar{u},v}(t,f) M_{u,\bar{v}}(t,f).
    \label{eq:M_uv}
\end{equation}

\subsection{Mask estimation using BLSTM}

To estimate the mask, the method first extracts the log absolute value $\mathbf{L}_{u,v} \in [0,+\infty]$ and the phase $\mathbf{P}_{u,v} \in [-\pi,+\pi]$ from the cross-spectrum $\mathbf{Y}_{u,v}$ as:
\begin{equation}
    \mathbf{L}_{u,v} = \log(\|\mathbf{Y}_{u,v}\|^2_2+\epsilon) - \log(\epsilon),
\end{equation}
\begin{equation}
    \mathbf{P}_{u,v} = \angle\mathbf{Y}_{u,v} ,
\end{equation}
where the constant $\epsilon$ holds a small value (here set to $10^{-20}$) to avoid large negative values as the energy goes to zero and $\angle$ stands for the angle.
Both features are then concatenated as:
\begin{equation}
    \mathbf{C}_{u,v} = (\mathbf{L}_{u,v},\mathbf{P}_{u,v}).
\end{equation}

\begin{figure}[!ht]
    \centering
    \includegraphics[width=\linewidth]{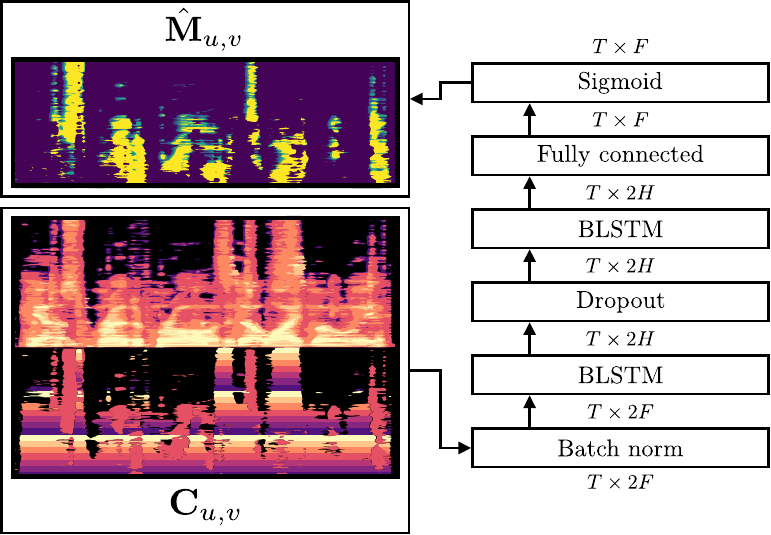}
    \vspace{-5pt}
    \caption{Architecture of the BLSTM network. The expressions $T$, $F$ and $H$ stand for the number of frames, the number of frequency bins and the number of hidden states, respectively.}
    \label{fig:lstm}
\end{figure}

The ideal mask $\mathbf{M}_{u,v} \in \mathcal{U}^{T \times F}$ introduced in (\ref{eq:M_uv}) is then estimated from $\mathbf{C}_{u,v} \in \mathbb{R}^{T \times 2F}$ using the following non-linear function:
\begin{equation}
    g: \mathbb{R}^{T \times 2F} \rightarrow \mathcal{U}^{T \times F},
\end{equation}
where the set $\mathcal{U} = [0,1]$ as the soft mask lies between $0$ and $1$, $T$ stands for the number of frames and $F$ for the number of frequency bins.
For this task, the method uses a BLSTM \cite{greff2016lstm} with two layers with a hidden size of $2H = 256$ and one dropout layer (with a probability of $p=0.2$), as shown in Fig. \ref{fig:lstm}.
A batch norm layer is also added to speed up convergence while training.
The BLSTM generates the estimated mask $\hat{\mathbf{M}}_{u,v}$ for each microphone pair at index $(u,v)$:
\begin{equation}
    \hat{\mathbf{M}}_{u,v} = g(\mathbf{C}_{u,v}).
\end{equation}

During training, the loss function $L$ corresponds to the mean square error weighted by the log absolute value of the cross-spectrum to give more weight to time-frequency regions dominated by speech \cite{wang2018supervised} and ignore silence periods:
\begin{equation}
    L = \|(\mathbf{M}_{u,v} - \hat{\mathbf{M}}_{u,v}) \odot \mathbf{L}_{u,v}\|^2_2,
\end{equation}
where $\odot$ stands for the Hadamard product.

Once the BLSTM is trained, it is used to estimate the pairwise masks by inference, and the overall mask is obtained according to:
\begin{equation}
    \hat{\mathbf{M}} = \frac{1}{|\mathcal{Q}|}\sum_{(u,v) \in \mathcal{Q}}{\hat{\mathbf{M}}_{u,v}},
\end{equation}
where $|\dots|$ stands for the cardinality of the set.

\subsection{GEV-BAN beamforming}

As suggested in \cite{heymann2015blstm}, the target and noise covariance matrices can be estimated with a soft mask $M_{\nu}$ between $0$ and $1$:
\begin{equation}
    \bm{\Phi}_{\nu\nu}(f) = \sum_{t \in \mathcal{T}}{M_{\nu}(t,f)\mathbf{Y}(t,f) \mathbf{Y}(t,f)^H},
\end{equation}
where $\mathcal{T} = \{1,\dots,T\}$, $\nu \in \{\mathbf{X},\mathbf{N}\}$ (with $\mathbf{X}$ being the target, and $\mathbf{N}$ being interference and background noise), $\mathbf{Y}(t,f) \in \mathbb{C}^{M \times 1}$, $\{\dots\}^H$ is the Hermitian transpose and:
\begin{equation}
    M_{\nu}(t,f) = 
    \begin{cases}
        \hat{M}(t,f) & \nu = \mathbf{X} \\
        1 - \hat{M}(t,f) & \nu = \mathbf{N} \\
    \end{cases}.
\end{equation}

The vector $\mathbf{F}_{GEV}(f) \in \mathbb{C}^M$ then corresponds to the principal component of the following generalized eigenvalue decomposition:
\begin{equation}
    \mathbf{F}_{GEV}(f) = \mathcal{P}\{\bm{\Phi}_{\mathbf{NN}}(f)^{-1}\bm{\Phi}_{\mathbf{XX}}(f)\},
\end{equation}
where $\mathcal{P}\{\dots\}$ stands for the principal component and $\{\dots\}^{-1}$ for the matrix inverse.
Heymann et al. \cite{heymann2015blstm} also suggest using a blind analytic normalization (BAN) gain to cope with potential non-linear distortion of the target source, as follows:
\begin{equation}
    g_{BAN}(f) = \frac{\sqrt{\mathbf{F}^H_{GEV}(f)\bm{\Phi}_{\mathbf{NN}}(f)\bm{\Phi}_{\mathbf{NN}}(f)\mathbf{F}_{GEV}(f)}}{\mathbf{F}^H_{GEV}(f)\bm{\Phi}_{\mathbf{NN}}(f)\mathbf{F}_{GEV}(f)D^2}.
\end{equation}

Finally, the reconstructed spectrogram for the target source $Z(t,f)$ can be obtained according to:
\begin{equation}
    Z(t,f) = g_{BAN}(f) \mathbf{F}^H_{GEV}(f)\mathbf{Y}(t,f).
\end{equation}

\section{Dataset}
\label{sec:dataset}

To train the network, we generate a dataset of synthetic stereo speech mixtures in simulated reverberating rooms.
The speech segments for training come from the LibriSpeech ASR corpus \cite{pa2015librispeech} that contains 360 hours of English text read by 482 men and 439 women, sampled at $f_S = 16000$ samples/sec.
A simulator based on the image method \cite{habets2006room} generates $10,000$ room impulse responses (RIRs).

For the network to generalize to various conditions, we sample the parameters in Table \ref{tab:RIR parameters} according to a uniform distribution.
Each RIR is defined by the room dimensions, reflection coefficient and speed of sound.
The spacing between both microphones varies to generalize to arbitrary microphone array shapes, and the microphone pair is rotated randomly and positioned in the room by making sure there is a minimum distance between the microphones and all surfaces.
Moreover, the sources are positioned randomly in the room in such a way that the distance between them and the microphones lie within a defined range.

For each training sample, we convolve two speech segments of 5 seconds from Librispeech (one for the target and the other one for the interference) with one of the generated RIR.
A randomly selected signal-to-noise ratio (SNR) then defines the gain of each source.
A random gain is also applied to each microphone, to cope with the potential gain mismatch between the microphones.
Some diffuse white noise with random variance is then added to the mixture.
Finally, all the signals are scaled by a common linear gain such that the signal range models scenarios with different volume levels.
The STFT uses frames of $N = 512$ samples, spaced by $\Delta N = 128$ samples.

\begin{table}[!ht]
    \centering
    \caption{Simulation parameters.}
    \vspace{-8pt}
    \def\arraystretch{1.3}
    \begin{tabular}{|cc|}
        \hline
        Parameters & Range \\
        \hline
        Room length (m) & $[5.0,10.0]$ \\
        Room width (m) & $[5.0,10.0]$ \\
        Room height (m) & $[2.0,5.0]$ \\
        Surfaces reflection coefficient & $[0.2, 0.8]$ \\
        Speed of sound (m/s) & $[340.0, 355.0]$ \\
        Spacing between mics (m) & $[0.04, 0.20]$ \\
        Min. dist. between mics and surfaces (m) & $0.5$\\
        Dist. between sources and mics (m) & $[1.0, 5.0]$\\
        White noise variance & $[0.5,2.0]$\\
        Signal to noise ratio (dB) & $[-5.0,+5.0]$ \\
        Overall linear gain & $[0.01,0.99]$\\
        \hline
    \end{tabular}
    \label{tab:RIR parameters}
\end{table}

\begin{figure}[!ht]
    \centering
    \subfloat[ReSpeaker USB\cite{sudharsan2019ai}]{\includegraphics[width=0.33\linewidth]{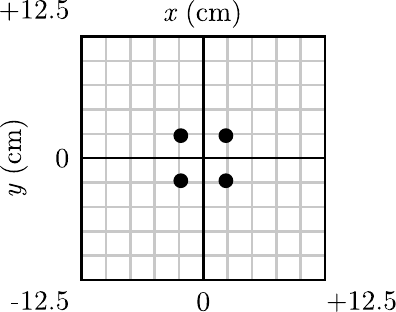}}
    \subfloat[ReSpeaker Core]{\includegraphics[width=0.33\linewidth]{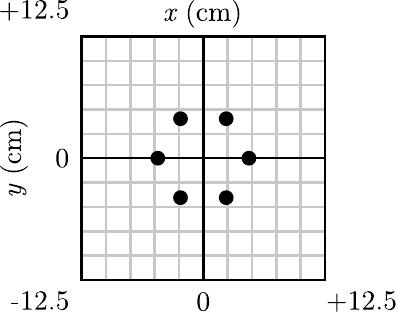}}
    \subfloat[Matrix Creator \cite{haider2019system}]{\includegraphics[width=0.33\linewidth]{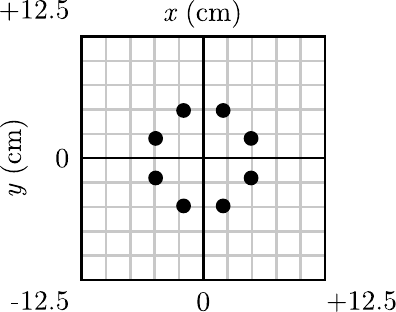}}\\\vspace{-5pt}
    \subfloat[Matrix Voice]{\includegraphics[width=0.33\linewidth]{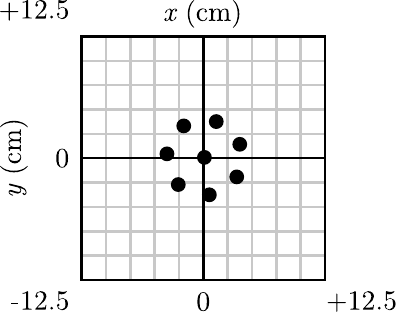}}
    \subfloat[MiniDSP UMA \cite{agarwal2018opportunistic}]{\includegraphics[width=0.33\linewidth]{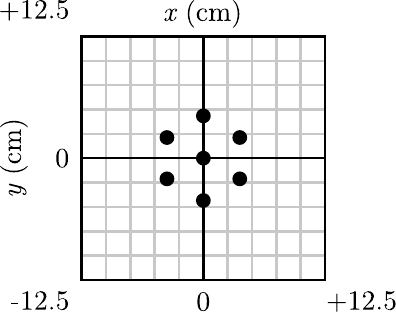}}
    \subfloat[MS Kinect \cite{pei2013sound}]{\includegraphics[width=0.33\linewidth]{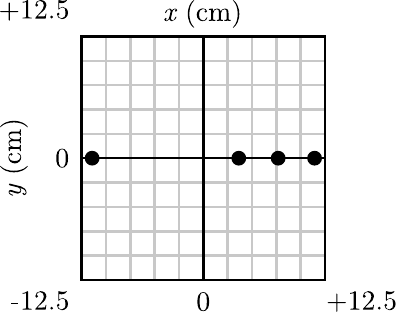}}
    \vspace{-5pt}
    \caption{Microphone array geometries}
    \label{fig:micarray_geometry}
\end{figure}

At test time, we use the test set from LibriSpeech, and convolve the sound segments with $1000$ RIRs generated for each array geometry.
We use the same simulation parameters as in Table \ref{tab:RIR parameters}, but ignore the spacing between microphones as the shapes correspond to the geometries of commercially available microphone arrays, as depicted in Figure \ref{fig:micarray_geometry}.
All these microphone arrays are planar, which means they span the $xy$-plane.
Note that the target and interference sources are positioned such that there is at least one pair of microphones that leads to discriminative TDOAs.

\section{Results and Discussion}
\label{sec:results}

\begin{figure}[!ht]
    \centering
    \subfloat[Reference signal at Microphone 1]{\includegraphics[width=\linewidth]{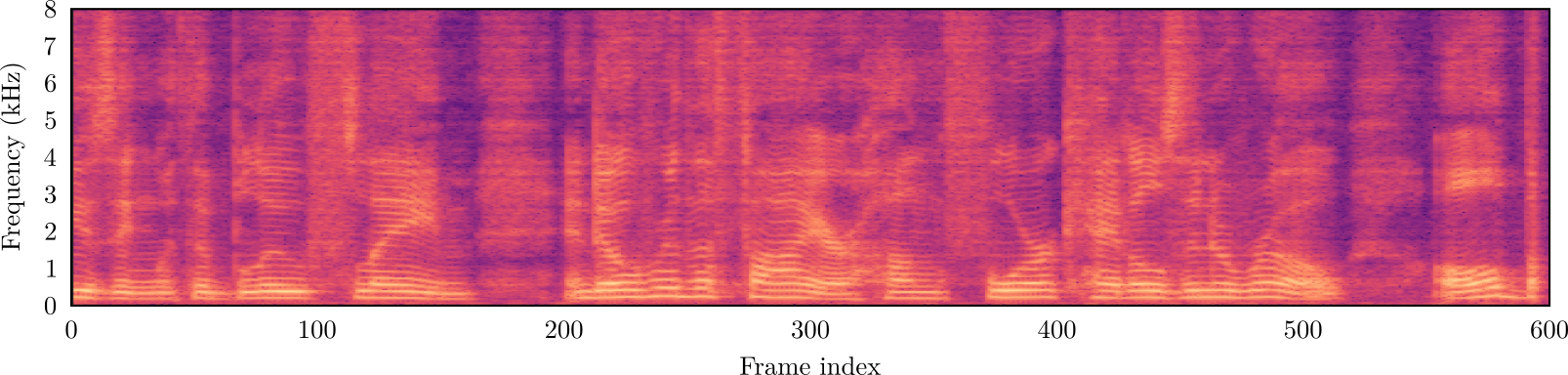}}\vspace{-8pt}\\
    \subfloat[Mixture at Microphone 1 (SDR = -6 dB)]{\includegraphics[width=\linewidth]{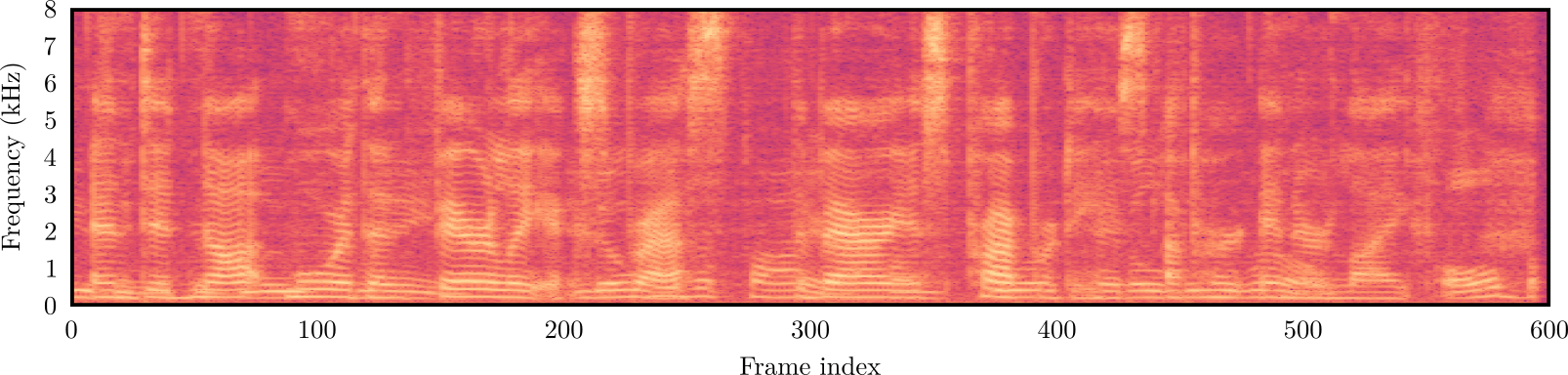}}\vspace{-8pt}\\    
    \subfloat[Separated signal with SteerNet (SDR = +9 dB)]{\includegraphics[width=\linewidth]{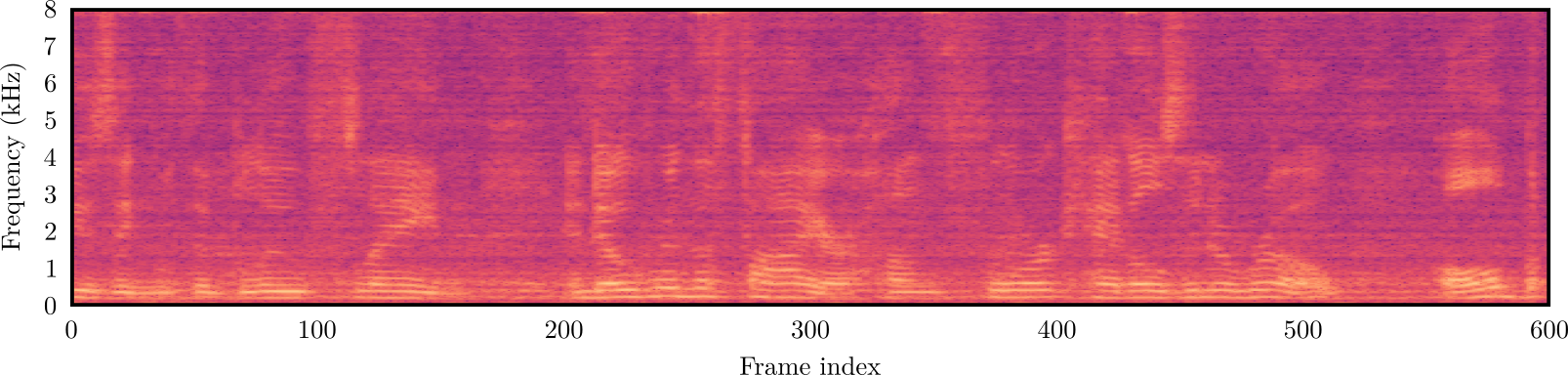}}\vspace{-5pt}    
    \caption{Example with the Matrix Voice 8-microphone array.}
    \label{fig:matrix_voice}
\end{figure}

Results demonstrate that SteerNet performs efficient separation for a wide range of microphone array geometries and environmental conditions.
For example, Fig. \ref{fig:matrix_voice} shows the reference signal, the mixture and the separated signal with a Matrix Voice microphone array.
Most features (formants, pitch, transient, etc.) are properly restored without any non-linear distortion, which is expected with GEV-BAN beamforming.
However, there is some extra energy in the low frequencies, which is also expected as the Matrix Voice microphone array has a small aperture, making separation more challenging in low frequencies.

The Signal-to-Distortion Ratio (SDR) is also computed for all the test samples using the BSS Eval toolbox \cite{vincent2006performance}. 
Table \ref{tab:sdr} shows that the SDR improves with all microphone arrays, regardless of the shapes.
This confirms that SteerNet enhances a target speech signal using its DOA and a trained BLSTM that generalizes for any pairs of microphones.
It should be noted that the network is trained using the Adam optimizer with a learning rate of $0.001$ and converges in around 20 epochs.
The parameters to estimate the gain for the oracle mask during training in (\ref{eq:G}) are set to $\alpha=10.0$ and $\beta=1.0$.
The Python code with audio samples is available online\footnote{https://github.com/francoisgrondin/steernet}.

\begin{table}[!ht]
    \centering
    \caption{SDR improvement (more is better).}
    \vspace{-8pt}
    \def\arraystretch{1.3}
    \begin{tabular}{|cc|}
        \hline
        Microphone Array & $\Delta$SDR (dB) \\
        \hline
        ReSpeaker USB & +7.69 \\
        ReSpeaker Core & +5.63 \\
        Matrix Creator & +5.13 \\
        Matrix Voice & +4.78 \\
        MiniDSP UMA & +4.78 \\
        Microsoft Kinect & +6.83 \\
        \hline
    \end{tabular}
    \label{tab:sdr}
\end{table}

The next step would be to optimize the hyperparameters for the proposed BLSTM architecture, and also investigate other neural network architectures.
Moreover, SteerNet considers only one interfering source.
This number could be increased to reflect more complex interaction scenarios.
Background noise from various environments could also make the model more representative of real-life scenarios.
Finally, it would be relevant to reduce the time context (currently set to 5 seconds) to adapt the approach to online processing with low latency.

\clearpage
\bibliographystyle{IEEEtran}

\bibliography{mybib}

\end{document}